\documentclass[aps,prd,
groupedaddress,showpacs,nofootinbib,amssymb,balancelastpage,preprintnumbers]{revtex4}

\usepackage[dvipdfmx]{graphicx}
\usepackage{graphicx,bm,color}

\usepackage{amsmath}
\usepackage{amssymb}
\usepackage{amsfonts}
\usepackage{cases}
\usepackage{cancel}
\usepackage{hyperref}
\usepackage{ulem}
\usepackage{subfigure}
\usepackage{here}
\usepackage{color}
\usepackage{tikz}
\usetikzlibrary{matrix}

\begin{document}

\title{Axial Inverse Magnetic Catalysis }

\author{Yuanyuan Wang}\thanks{{\tt yuanyuanw20@mails.jlu.edu.cn}}
\affiliation{Center for Theoretical Physics and College of Physics, Jilin University, Changchun, 130012,
China}

\author{Shinya Matsuzaki}\thanks{{\tt synya@jlu.edu.cn}}
\affiliation{Center for Theoretical Physics and College of Physics, Jilin University, Changchun, 130012,
China}

\begin{abstract}

We find that the inverse magnetic catalysis for $U(1)$ axial symmetry (AIMC: axial inverse magnetic catalysis) can emerge around 
the chiral crossover regime in the thermomagnetic QCD with 2 + 1 flavors 
at physical point. 
This phenomenon can be correlated with the IMC for the chiral $SU(2)_L \times SU(2)_R$ symmetry (CIMC: chiral IMC). 
We explicitly observe the AIMC based on a Nambu-Jona-Lasinio model with 2 + 1 quark flavors, 
where introduced anomalous magnetic moments of the quarks play the essential role to 
drive both the CIMC and AIMC. 
Our finding is shortly testable on lattices. 
Possible phenomenological and 
cosmological implications are also briefly addressed.

\end{abstract}

\maketitle

\section{Introduction}

Violation of $U(1)$ axial symmetry plays a key role to address the QCD vacuum 
characterized by quark condensates, as well as the chiral symmetry breaking.
In particular, it has been a longstanding issue how much the $U(1)_A$ 
breaking contributes to the quark condensate. This question could be 
related to what is the major role for the origin of mass in the thermal history of 
the universe.

The state-of-art-lattice simulations have so far clarified 
that in the hot QCD with 2 + 1 (light up and down 
quarks and one heavy strange quark), 
the $U(1)_A$ axial breaking tends to survive 
longer than the chiral $SU(2)$ breaking for the lightest two flavors as 
temperature grows~\cite{Aoki:2012yj,Bhattacharya:2014ara}. 
This result is also supported from 
a rigorous argument based on QCD-inequality like 
relations~\cite{Cohen:1996ng} and 
its generalized evidence based on the lattice QCD setup~\cite{Cohen:1997hz}. 
Moreover, a recent lattice study (with two lightest flavors)  
has shown a hint that significantly 
dominant contributions from the $U(1)_A$ breaking  
are left in the quark condensate, 
during the chiral phase transition (crossover)~\cite{Aoki:2021qws}. 
All those may imply that in a view of the thermal history of the universe, 
the main source for the origin of mass is 
supplied from the $U(1)_A$ breaking.

However, it might be not the end of the story: 
even in early universe including the QCD phase transition epoch, 
a strong enough magnetic field might be present as a background field, which 
could be generated due to some primordial (electroweak) phase transitions of strong first order~\cite{Vachaspati:1991nm,Enqvist:1993np,Grasso:1997nx,Grasso:2000wj,Ellis:2019tjf,Di:2020ivg,Yang:2021uid}. Therefore, the QCD dynamics in the thermal history might be thermomagnetic, which is called thermomagnetic QCD. 
In addition, 
heavy ion collision experiments~\cite{Kharzeev:2007jp,Skokov:2009qp,Deng:2012pc,Bloczynski:2013mca,Hirono:2012rt,Deng:2014uja,Voronyuk:2014rna,Huang:2015oca} can create 
a strong enough magnetic field in the thermal plasma of QCD, 
due to the relativistic motion of the colliding nuclei and the smallness of the system. 
Thus, the thermomagnetic QCD has nowadays opened a vast ballpark, 
non only with cosmological, but also experimental interests, 
involving lattice simulations, and chiral-effective model approaches.

Some outstanding results have already been reported from 
lattice studies on the QCD thermodynamics in a strong external magnetic field~ \cite{DElia:2012ems,Endrodi:2014vza}.  
Of particular importance related to the chiral symmetry breaking are the reduction of the pseudo-critical temperature for the chiral  crossover~\cite{Bornyakov:2013eya, Bali:2014kia, Tomiya:2019nym, DElia:2018xwo, Endrodi:2019zrl}
and the inverse magnetic catalysis~\cite{Bali:2011qj}. 
Both two imply a faster effective restoration of the chiral symmetry 
in hot-magnetized early universe.  
To our best knowledge, however, no definite analysis on the $U(1)_A$ breaking and effective 
restoration in the thermomagnetic QCD has been curried out, or no implications of 
the inverse magnetic catalysis for the chiral symmetry to the 
$U(1)_A$ symmetry has been argued.

In this paper, prior to lattice simulations in the near future, 
based on a chiral effective model we observe a definite implication 
of the inverse magnetic catalysis for the chiral symmetry, 
to the $U(1)_A$ breaking in thermomagnetized QCD: 
that is the inverse magnetic catalysis (IMC) for the $U(1)_A$ symmetry, 
which we call the axial IMC (AIMC), and in comparison, we call 
the IMC for the chiral $SU(2)_L \times SU(2)_R$ symmetry CIMC. 
We employ a Nambu–Jona-Lasinio (NJL) model with 2 + 1 flavors at the 
physical point, 
in which we introduce a couple of anomalous magnetic moments (AMMs) of quarks, 
as well as so-called Kobayashi-Maskawa-`t Hooft determinant term~\cite{Kobayashi:1970ji,Kobayashi:1971qz,tHooft:1976rip,tHooft:1976snw} mimicked as the instanton-induced $U(1)_A$ anomaly in the underlying QCD.

We first look into the viable parameter space to realize the CIMC, 
which is mapped on the AMM parameters. Then we evaluate 
the difference of susceptibilities for the $U(1)_A$ partner, 
such as $\pi$ and $\delta$ mesons, denoted as $\chi_\pi$ and $\chi_\delta$ 
as a function of the applied constant magnetic field and temperature. 
In terms of the axial susceptibility, $\chi_{\pi- \delta} \equiv \chi_\pi - \chi_\delta$, the AIMC is dictated by 
observing that at zero temperature, 
$\chi_{\pi - \delta}$ becomes larger and larger, as the magnetic field strength 
increases, and  
the pseudo-critical temperature of $\chi_{\pi- \delta}$, $T_{pc}^A$, 
gets smaller as the magnetic field gets stronger. 
Here $T_{pc}^A$ is defined as 
the inflection point of the temperature evolution of $\chi_{\pi- \delta}$.

In light of the CIMC, the two-flavor NJL model with AMMs of quarks 
has so far been discussed~\cite{Ferrer:2014qka,Fayazbakhsh:2014mca,Chaudhuri:2019lbw,Chaudhuri:2020lga,Ghosh:2020xwp,Xu:2020yag,Farias:2021fci}. 
There, in terms of the symmetry argument, 
realization of the CIMC can be understood by emergence of 
destructive interference in the light quark condensates 
between two explicit chiral-breaking sources, where 
one comes from the current quark mass, while the other from the AMMs of quarks 
coupled to a strong magnetic field. 
No explicit work on the chiral phase transition has been done with 2 + 1 flavors coupled with the AMMs in the framework of NJL.

We will shed the first light on the 2 + 1 flavor case, 
and clarify the fully viable parameter space to realize the desired CIMC at 
high temperatures, and MC at zero temperature. 
We find that at any temperature including zero temperature, 
the AMMs of $u$ and $d$ quarks contribute to 
the $u$ and $d$ quark condensates, or the constituent-$u$ and $d$ quark masses, destructively 
against effects from the current quark mass and the $U(1)_A$ anomaly, 
while the AMM of strange quark  
acts as destructive interference at lower temperature, and constructive one at 
higher temperatures: namely, the AMM of strange quark 
tends to cease realization of MC and CIMC, at zero temperature and higher temperatures, respectively. This latter feature is a new finding 
characteristic to the 2 + 1 flavor model.

Since the AMM interactions break not merely the chiral $SU(2)_L \times SU(2)_R$ symmetry, but the $U(1)_A$ symmetry,  similar destructive interference is expected to happen in the $U(1)_A$ sector, i.e., $\chi_{\pi - \delta}$.
Such coincidental correlation between the chiral and $U(1)_A$ symmetries  may also be deduced in a context of the QCD-inequality argument~\cite{Cohen:1996ng,Aoki:2012yj}, and a recent lattice study detailed on the Dirac spectrum~\cite{Ding:2020xlj,Tomiya:2016jwr}. 
The latter, in particular, works through an operator identity-like relation
between the Dirac spectrum, the chiral condensate, and the $U(1)_A$ susceptibility, and should hold even including external fields, such as a constant magnetic field~\footnote{
We thank Akio Tomiya for comments on the validity of identities with the Dirac spectrum. 
}. 
Still, however, it would be nontrivial in a sense of framework of chiral effective models  
to monitor such chira- $U(1)_A$ axial coherence in the 2 + 1 flavor case.  
Based on the NJL model with AMMs, we will explicitly clarify how efficiently the destructive interference arises in the $U(1)_A$ susceptibility $\chi_{\pi -\delta}$, coherently in the chiral condensate.

We find that the AIMC can take place in the chiral crossover regime where the CIMC is present, which depends on the size of AMM, and the pseudo-critical temperature $T_{pc}^A$ drops as increasing the magnetic field 
strength. 
Lattice simulations in the near future will give 
a decisive conclusion on existence of the AIMC and the competition for the survival between the chiral and $U(1)_A$ breaking at higher temperatures.

This AIMC may provide a hint to reveal whether  
in a sense of early thermomagnetic universe, 
the remnant of the $U(1)_A$ breaking in the origin of mass 
might be comparable with what the chiral breaking leaves, 
in contrast to the pure-thermal QCD in which the former might highly dominate~\cite{Aoki:2021qws}.

Other phenomenological and cosmological implications deduced 
from the AIMC are also briefly addressed.



\section{NJL model with quark AMM: preliminaries}  

To clarify the presence of AIMC, 
we 
work on an NJL model with 2 + 1 flavors  
with AMMs of quarks in a constant magnetic field. 
The model is described by the following Lagrangian:  
\begin{equation}
	\begin{aligned}
		\mathcal{L}=
	  \sum_{f=u,d,s} 
		\bar{\psi}_f\left(i \gamma^{\mu} D_{\mu}^{(f)}-\hat{m}_{0f}+\kappa_{f}(eB,T) q_{f} F_{\mu \nu} \sigma^{\mu \nu}\right) \psi_f 
		&+ G\left\{(\bar{\psi} \lambda^{a}\psi)^{2}+\left(\bar{\psi} i \gamma^{5} \lambda^{a} \psi\right)^{2}\right\}\\&-K\left[\operatorname{det} \bar{\psi}\left(1+\gamma_{5}\right) \psi+\operatorname{det} \bar{\psi}\left(1-\gamma_{5}\right) \psi\right].
	\end{aligned}
\label{Lag:NJL}
\end{equation}
Here $\psi_f=(u,d,s)^{T}$ denotes 
the three-flavor quark field forming the $SU(3)$ triplet;  
for simplicity, the current-quark mass matrix $\hat{m}_{0f}$ is taken to be diagonal in the flavor space: $\hat{m}_{0f}={\rm diag}(m_{0u},m_{0d},m_{0s})$, where the present 2 + 1 flavor setup leads to $m_{0u} = m_{od} \equiv m_0$; 
$\lambda^a$ ($a=0, \cdots , 8$) are the Gell-Mann matrices in the flavor space with $\lambda^0= \sqrt{2/3}$ ${\rm diag}(1,1,1)$; 
the covariant derivative $D^{(f)}_{\mu}=\partial_{\mu}-iq_{f}A_{\mu}$ contains the coupling between the quark and the external magnetic field with the electromagnetic charge matrix $q_f = e \cdot {\rm diag}\{ 2/3, -1/3, -1/3 \}$, where the magnetic field is applied along 
the $z$ direction and embedded in the electromagnetic gauge field $A_\mu$ as   
$A_{\mu}=(0,0,Bx,0 )$;  
$F_{\mu\nu}=\partial_{\mu} A_\nu -\partial_\nu A_{\mu}$ is the electromagnetic field strength; $G$ and $K$ are the four-fermion coupling constant and the determinant (six-fermion) coupling constant, respectively; $\sigma^{\mu\nu}=\frac{i}{2}\left[\gamma^{\mu},\gamma^{\nu}\right]$.  
The $\kappa_f(eB,T)$ is the AMM coupling, which is related to the AMM for quark $a_f$ as $a_f = 4 M_f \cdot \kappa_f$ with the constituent quark mass $M_f$. 
Here $\kappa_f$ itself is dependent of quark flavors, the background temperature $T$, and 
an applied strong enough magnetic field $B$, to be fixed later (See Eq.(\ref{kappa-v-sigma})).

Under the chiral $U(3)_L \times U(3)_R$ transformation: $\psi \to U \cdot \psi$ with $U= \exp[ - i \gamma_5 \sum_{a=0}^8  (\lambda^a/2) \theta^a ]$ and the chiral phases $\theta^a$, 
the four-fermion  interaction  term 
is $U(3)_L\times U(3)_R$ invariant. 
The mass term 
explicitly breaks the chiral $U(3)_L\times U(3)_R$ symmetry. 
The determinant term, called Kobayashi-Maskawa-`t Hooft (KMT) determinant~\cite{Kobayashi:1970ji,Kobayashi:1971qz,tHooft:1976rip,tHooft:1976snw}, 
is induced from the instanton coupled to quarks in the underlying QCD,  
and preserves the chiral $SU(3)_L\times SU(3)_R$ invariance (associated with the chiral phases labeled as $a=1, \cdots , 8$), but breaks the $U(1)_A$ (corresponding to $a=0$) symmetry. 
The AMM term with the coupling $\kappa_f$ 
explicitly breaks not only the chiral $SU(3)_L\times SU(3)_R$ symmetry, but also 
the $U(1)_A$ axial symmetry.


In addition to the quark mass terms, the KMT determinant term, and the AMM term, 
the chiral $U(3)_L\times U(3)_R$ symmetry is spontaneously broken 
by the nonperturbative dynamics of the present NJL. 
To monitor the spontaneous breaking, 
we simply employ the mean-field approximation, and construct 
the thermodynamic potential in the presence of a constant magnetic field. 
The thermodynamic potential is then 
given as a function of thermally averaged quark condensates $\langle \bar{u}u \rangle, \langle \bar{d}d \rangle$, and $\langle \bar{s}s \rangle$, 
are determined via the stationary condition of the potential, i.e., the gap equations:   
$\langle \bar{f}f \rangle = - i N_c \operatorname{tr} \int\frac{d^4p}{(2\pi)^4}S^{f}(p)$, 
where $S^f(p)$ stands for the full propagator of $f$-quark and $N_c$ is the number of 
colors, to be fixed to three. 
With respect to those quark condensates, the thermodynamic potential 
is minimized at the nontrivial vacuum with the stationary condition. 
We then find the coupled gap equations in terms of the constituent quark masses 
$M_f=(M_u, M_d, M_s)$, 
which take the same form as in the case without the magnetic field~\cite{Rehberg:1995kh}:   
\begin{align} 
&M_{f}=m_{0f}+\sigma_{f} 
\,, \label{def:sigmaf}
\end{align} 
where 
\begin{align} 
\sigma_u &= 4i G\langle \bar{u}u \rangle 
- 2K \langle \bar{d}d \rangle \langle \bar{s}s \rangle
\,, \notag \\
\sigma_d &= 4i G\langle \bar{d}d \rangle 
- 2K \langle \bar{u}u \rangle \langle \bar{s}s \rangle
\,, \notag \\
\sigma_s &= 4i G\langle \bar{s}s \rangle 
- 2K \langle \bar{u}u \rangle \langle \bar{d}d \rangle
\,. \label{gap:eqs}
\end{align}


In evaluating the thermally averaged quark condensates, 
we apply the imaginary time formalism. 
Taking into account nonzero constant magnetic field 
applied along the $z$-direction as well, 
we shall make the following replacements:  
\begin{align} 
p_{0} &\rightarrow i \omega_{\bf k}=i(2 {\bf k}+1) \pi T \,, \notag \\
\int \frac{d^{4} p}{(2 \pi)^{4}} &\rightarrow i T \sum_{{\bf k}=-\infty}^{\infty} \int \frac{d^{3} p}{(2 \pi)^{3}}
\to 
 i T \sum_{{\bf k}=-\infty}^{\infty} \sum_{f = u,d,s}  
 \sum_{n=-\infty}^{\infty} \frac{|q_f B|}{2\pi} 
 \int_{- \infty}^\infty \frac{d p_3}{2 \pi}
,  
\label{dp}
\end{align}
with the Matsubara frequency $\omega_{\bf k}$ and the Landau level $n$. 
Thus the analysis will be straightforward just by extending 
the one in the literature~\cite{Rehberg:1995kh} to the case with the 
magnetic field and AMMs.  


The $f$-quark propagator $S^{f}(p)$ including the AMM term 
is available in the literature 
~\cite{Xu:2020yag}, which , 
takes the form 
\begin{equation}
	\begin{aligned}
	S^{f}(p) &=i e^{-\frac{\vec{p}_{\perp}^{2}}{\left|q_{f} B\right|}} \sum_{n=0}^{\infty} \frac{D_{n}\left(q_{f} B, p\right) F_{n}\left(q_{f} B, p\right)}{A_{n}\left(q_{f} B, p\right)} \,, 
	\end{aligned}
	\label{Sf}
\end{equation}
where $p_\perp= (p_1, p_2)$, and~\footnote{
Here $L_n^\alpha$ denotes the generalized Laguerre polynomials. 
} 
\begin{equation}
	\begin{aligned}
	\\D_{n}\left(q_{f} B, p\right) &=\left(p^{0} \gamma^{0}-p^{3} \gamma^{3}+M_{f}+\kappa_{f} q_{f} B \sigma^{12}\right)\\&\times\left[\left(1+i \gamma^{1} \gamma^{2} \operatorname{sign}\left(q_{f} B\right)\right) L_{n}\left(\frac{2 \vec{p}_{\perp}^{2}}{\left|q_{f} B\right|}\right)-\left(1-i \gamma^{1} \gamma^{2} \operatorname{sign}\left(q_{f} B\right)\right) L_{n-1}\left(\frac{2 \vec{p}_{\perp}^{2}}{\left|q_{f} B\right|}\right)\right]\\& 
+4\left(p^{1} \gamma^{1}+p^{2} \gamma^{2}\right) L_{n-1}^{1}\left(\frac{2 \vec{p}_{\perp}^{2}}{\left|q_{f} B\right|}\right) \, ,
	\end{aligned}
\end{equation}
\begin{equation}
	\begin{aligned}
	F_{n}\left(q_{f} B, p\right) &=\left(\kappa_{f} q_{f} B-p^{0} \gamma^{3} \gamma^{5}+p^{3} \gamma^{0} \gamma^{5}\right)^{2}-M_{f}^{2}-2 n\left|q_{f} B\right| \,, 
	\end{aligned}
\end{equation}
\begin{equation}
	\begin{aligned}
A_{n}\left(q_{f} B, p\right) &=\left[\left(\kappa_{f} q_{f} B 
+ \sqrt{ p_{||}^2  }\right)^{2}-M_{f}^{2}-2 n\left|q_{f} B\right|\right] \\&\times\left[\left(\kappa_{f} q_{f} B-\sqrt{p_{||}^2} \right)^{2}-M_{f}^{2}-2 n\left|q_{f} B\right|\right] \, ,
	\end{aligned}
\end{equation} 
with $p_{||}^2 = p_0^2 - p_3^2$.

To regularize the intrinsic divergence terms arising  
in the momentum integration along the magnetic field direction ($p_3$), 
we adopt 
a smooth cutoff scheme 
with the cutoff function, 
\begin{equation}
	\begin{aligned}
	f_{\Lambda}(p_{3},n)=\frac{\Lambda^{10}}{\Lambda^{10}+(p^2_{3}+2n\lvert q_{f}B\rvert)^2}
	\,, 
	\end{aligned}
\end{equation}
where $\Lambda$ is the cutoff for the three-dimensional momentum integral 
in Eq.(\ref{dp}).

The AMM term breaks the full $U(3)_L \times U(3)_R$ 
symmetry even at the classical or perturbative level of QED. 
In a strong enough magnetic field, moreover, 
the AMM can actually be dynamically developed~\cite{Bicudo:1998qb,Ferrer:2008dy,Ferrer:2009nq,Chang:2010hb,Ferrer:2013noa,Ferrer:2014qka,Mao:2018jdo} 
as another chiral order parameter, simultaneously with the quark condensates.  
This dynamical generation would happen when the scale of the magnetic field strength would be 
comparable with or greater than the intrinsic infrared scale of QCD or the NJL dynamics, say the renormalization-group invariant $\Lambda_{\rm QCD}$, or (equivalently) 
dynamical quark mass at vacuum ($\sim 300$ MeV). 
Such strong enough magnetic field would cause 
the dimensional reduction: $D=4 \Rightarrow 2$ with 
the reduced Lorentz symmetry left in $D=2$: $SO(1,3) \Rightarrow SO(1, 1)  \times O(2)$, 
in which both the dynamics and kinematics are separately characterized by spaces parallel ($||$) and transverse ($\perp$) to 
the applied magnetic field direction (the $z$-direction). 
This is how the AMM operator $\bar{q}_f \sigma_{12} q_f$ would be allowed to condense 
in the residual-$SO(1,1) \times O(2)$ invariant way, to contribute to the dynamical chiral 
symmetry breaking, together with 
the normal quark condensate operator $\bar{q}_f q_f$, as has been 
discussed in the literature~\cite{Bicudo:1998qb,Ferrer:2008dy,Ferrer:2009nq,Chang:2010hb,Ferrer:2013noa,Ferrer:2014qka,Mao:2018jdo}.

In the present analysis, we assume the dynamical AMM to dominate over 
the perturbative contribution generated at the leading order of QED. 
Then, the AMM should be related to the dynamical mass part of the constituent quark mass. 
Inspired by the dynamical AMM generation via the NJL dynamics as in the literature listed above, 
but, still allowing 
the AMMs as free parameters~\footnote{
We have checked that the size of the NJL-driven AMM is too small to realize the CIMC at higher T. 
In this sense, it is necessary to go beyond 
the one-gluon exchange prescription
of the NJL for the AMM.  
 Note, however, that as long as we work on the leading order of the large $N_c$ expansion (or mean field approximation), even NJL-driven AMM term can be reduced to the $\kappa_f$ term in Eq.(\ref{Lag:NJL}). 
 Therefore, we can interpret the size of $v$ and $v_s$ as displayed in the plots later as the net contribution from the NJL-driven and beyond NJL effects.  
},   
we may model the relation between the AMMs ($\kappa_f$) and the dynamical mass parts of the quarks $(\sigma_f$ in Eq.(\ref{def:sigmaf})) as~\footnote{ 
A similar procedure has been 
applied in Ref.~\cite{Xu:2020yag} for the 2-flavor NJL case, with $v_u=v_d$ assumed. 
} 
\begin{align} 
\kappa_{f}(eB, T)= v_{f} \cdot \sigma_{f}(eB, T) \,,\qquad 
{\rm for } \quad eB \gtrsim eB|_{\rm min}
\,,  \label{kappa-v-sigma}
\end{align}
with the lower bound of the magnetic field strength, $\sqrt{eB}|_{\rm min}$, which is to be fixed later. 
The flavor-dependent coefficient $v_f$ is assumed to be flavor-universal for $u$ and $d$ 
quarks, i.e., $v_u = v_d \equiv v$, but is taken to be different for $s$ quark: $v_s \neq v$. 
This would be a reasonable setup because the AMM parameter $\kappa_f$ is defined 
exclusively out of the overall electromagnetic charge, as seen from Eq.(\ref{Lag:NJL}), 
and QCD (or NJL dynamics) is potentially flavor-universal, hence the discrepancy 
in AMMs among flavors arises only from the chiral-explicit breaking sources, which can be 
fully incorporated in the dynamical mass part $\sigma_f$.

In closing this preliminary section, 
we introduce five model parameters fixed at vacuum with $eB=T=0$:  
$m_{0}=0.0055$ GeV, $m_{0s}=0.1407$ GeV, $\Lambda=0.6023$ GeV, 
$G\Lambda^2=1.835$, and $K\Lambda^5=12.36$, 
which are quoted from the literature~\cite{Rehberg:1995kh},  
where inputs are an optimized and 
conventional set of the hadronic observables in the isospin-symmetric limit: 
$m_\pi = 0.135$ GeV, $m_K=0.4977$ GeV, $m_{\eta'}=0.9578$ GeV, 
$f_\pi = 0.0924$ GeV, and $m_0 = 5.5$ MeV.

We will not consider intrinsic-temperature dependent couplings, 
instead, all the $T$ dependence should be induced only 
from the thermal quark loop corrections to the couplings 
defined and introduced at vacuum. 
Actually, the present NJL at $eB=0$ shows good 
agreement with lattice QCD results on the temperature scaling 
for the chiral, axial, and topological susceptibilities, 
as shown in Ref.~\cite{Cui:2021bqf}. 
In this sense, we do not need to introduce such an 
intrinsic $T$ dependence for the model parameters 
in the regime up to temperatures around the 
chiral crossover.

\section{CIMC and ``phase" diagram} 
\begin{figure}[t] 
\centering
	\includegraphics[width=0.50\linewidth]{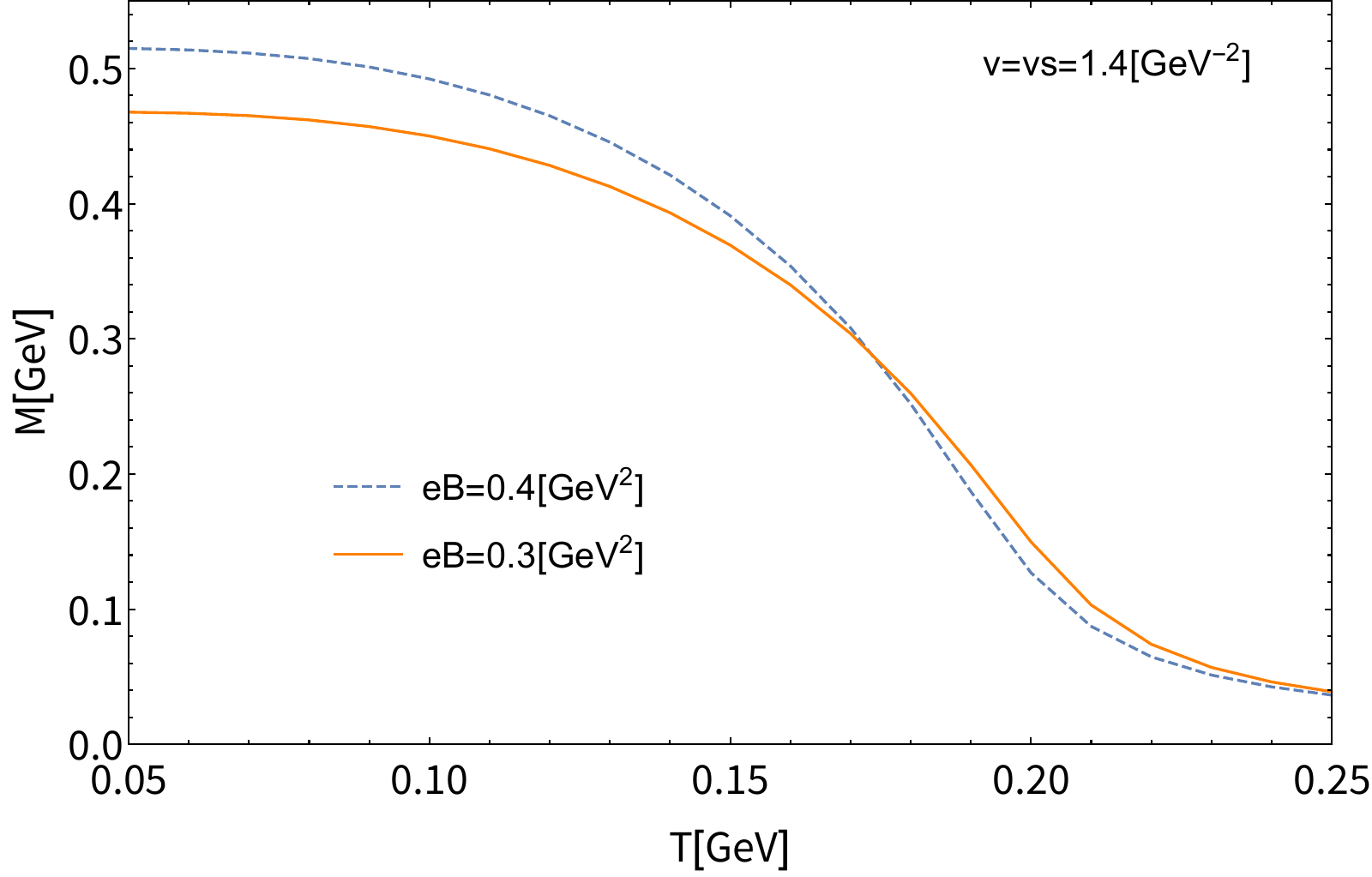}
	\caption{Plot of the averaged constituent mass for $u$ and $d$ quarks, $M$, versus 
	temperature evolution, with $eB$ varied and the AMM parameters $v$ and $v_s$ being 
	fixed to a reference point (included in the allowed regime of the ``phase" diagram in 
	Fig.~\ref{CIMC range11}).}  
\label{CIMC11}
\end{figure}

With the preliminary setup provided in the previous section, 
we first evaluate the constituent quark mass for $u$ and $d$ quarks 
as a function of temperature $T$, 
given the strength of the applied magnetic field and the AMM parameters $v$ and $v_s$. 
See Fig.~\ref{CIMC11}. We have taken $\sqrt{eB} \sim (0.5 - 0.6)$ GeV for 
a wide range of $T$, $(0.05 - 0.25)$ GeV, where the strength of magnetic field 
is strong enough in light of lattice QCD: 
the constant magnetic field applied on lattice QCD has the minimal size fixed as 
$|eB|_{\rm min} \propto T^2$~\cite{Bali:2011qj}, and one can check that 
$\sqrt{eB}|_{\rm min} < T_{\rm highest} \sim 0.25$ GeV $< \sqrt{{\rm eB}}|_{\rm applied} \sim 
(0.5- 0.6)$ GeV. 
In the figure, we have plotted the averaged mass for $u$ and $d$ quarks, $M=\frac{M_{u}+M_{d}}{2}$, which is a reasonable chiral order parameter 
in the presence of a couple of isospin breaking sources, 
and taken the AMM parameters $v = vs = 1.4\,{\rm GeV}^{-2}$, as a reference point to be clarified below. 
We see that for small $T$, $M$ increases as the magnetic field gets larger, 
while at higher $T$, it turns to decrease with the magnetic field. 
This confirms the CIMC phenomenon, and is a successful result generalized from 
the two-flavor NJL with the AMM~\cite{Ferrer:2014qka,Fayazbakhsh:2014mca,Chaudhuri:2019lbw,Chaudhuri:2020lga,Ghosh:2020xwp,Xu:2020yag,Farias:2021fci}, to the 2 + 1 flavor case including 
the KMT-determinant $U(1)_A$-anomaly contribution ($K$ terms in Eq.(\ref{gap:eqs})). 
The CIMC is successfully realized essentially due to a moderately large 
AMM for $u$ and $d$ quarks which constructively contributes to $M$ with 
the thermal corrections, to make $M$ dropped faster.

\begin{figure}[t]
	\centering
	\includegraphics[width=0.50\textwidth]{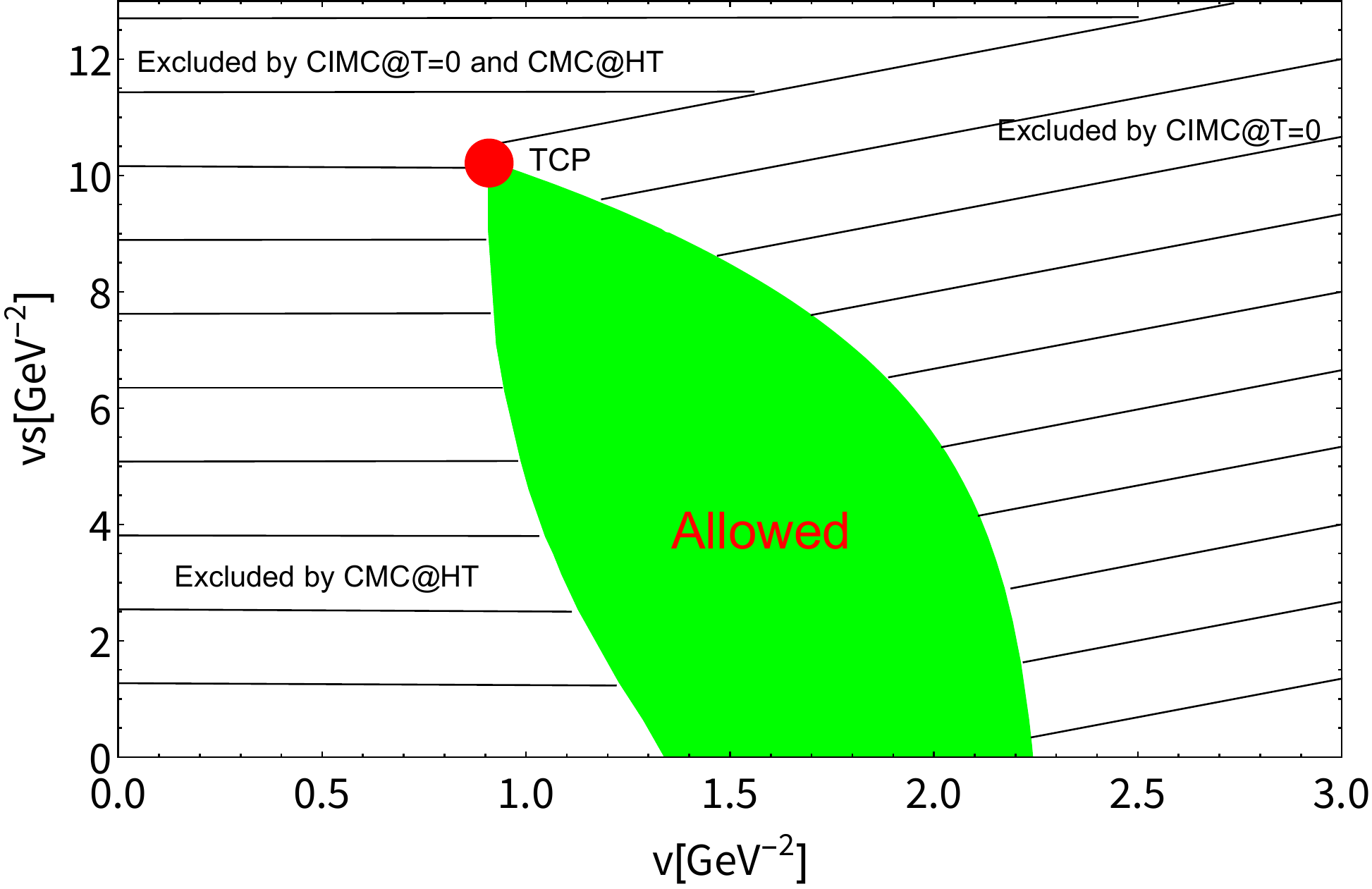}
	\caption{The ``phase" diagram on $(v, v_s)$ plane classified by the CMC and CIMC features 
	at $T=0$ and higher temperatures (HT). The three phases filled by shaded lines, corresponding to 
	[1], [2] and [3] in the text, are ruled out because of failure of realizing 
	the magnetic features on the chiral symmetry breaking, reported from lattice QCD. 
	Only the centered regime of green-leave shape is allowed. All the four phases 
	merge at the tetra-critical point, TCP, denoted as a blob. } 
	\label{CIMC range11}
\end{figure}


Scanning over the AMM parameter space $(v, v_s)$ with the size of $eB$ varied in an appropriately strong enough range (as noted above), we examine the MC and IMC features 
at $T=0$ and at higher $T$. 
Thus, the ``phase" diagram is drawn on the $(v, v_s)$ space, 
as depicted in Fig.~\ref{CIMC range11}. 
The diagram turns out to be divided into four ``phases", where the model realizes  
\begin{itemize} 

\item[[1]] 
CIMC for any $T$ including $T=0$, due to larger AMMs for $u$ and $d$ quarks;  

\item[[2]] 
CMC for any $T$ including $T=0$, due to smaller AMM for $u$ and $d$ quarks ;  

\item[[3]] 
CIMC at $T=0$, and CMC at higher temperatures, due to significant AMM for strange quark; 

\item[[4]] 
CMC at $T=0$, and CIMC at higher temperatures, due to moderate AMMs for 2 + 1 quarks. 
\end{itemize}
The first three ``phases" [1], [2], and [3] are excluded, because they do not 
reproduce the lattice results on the CMC and CIMC. 
Thus, only the ``phase" [4] survives, which corresponds to 
the ``Allowed" regime in Fig.~\ref{CIMC range11}. 
Once may notice that there is a critical point at which all four ``phases" 
merge on the diagram, that is, the tetra-critical point (TCP), 
which has been also specified in Fig.~\ref{CIMC range11}.

Of particular interest is to note the ``phase" [3], 
where the AMM of strange quark acts like a destructive interference 
against realization of the CMC at $T=0$ and CIMC at higher temperatures.  
This is in contrast to the role of the AMM for $u$ and $d$ quarks. 
It is operative in the light-quark constituent mass $(M)$, and  
contributes destructively against the current quark mass 
and the $U(1)_A$ anomaly at any temperature including zero temperature, 
so that when it is too large, the CIMC is driven even at $T=0$ (``phase" 1), 
while the CMC shows up at any temperature, 
when the light-quark AMM is too small (``phase" 2 ). 
This feature can also be observed by viewing the plot along the $v$ axis at $v_s=0$, in Fig.~\ref{CIMC range11}.

With the ``phase" diagram taken into account, 
below we will discuss the $T$- and $eB$- dependence on the axial susceptibility.


\section{Axial susceptibility: $\chi_{\pi - \delta}$, AMC, and AIMC} 

The axial susceptibility is constructed from difference of two 
susceptibilities related to the $U(1)_A$ partners. 
In the present NJL model with 2 + 1 flavors, 
we have two candidates of the $U(1)_A$ partners, which are, 
in terms of meson names, $(\sigma, \eta)$ and $(\pi, \delta)$. 
Those cases should be just alternatives each other, and exhibit the same axial 
property, which would be so even at finite $T$ and $eB$. 
In the present analysis, we take $\pi$ and $\delta$ meson channels, and 
investigation of the other partner is to be pursued in another publication. 

We start with evaluating the $\pi$ channel. 
The $\pi$ meson susceptibility $\chi_\pi$ is defined as 
\begin{equation}
	\chi_{\pi}
	=\int_T d^4 x 
	\left[ 
	\langle (\bar u(0)  i\gamma_5  u(0))( \bar u(x) i\gamma_5 u(x))\rangle_{\rm conn}
+ \langle ( \bar d(0)  i\gamma_5 d(0))(\bar d(x) i\gamma_5 d(x))\rangle_{\rm conn}
\right]
\, , \label{chi-pi:def}
\end{equation}
with $\langle \cdot \cdot \cdot \rangle_{\rm conn}$ being the connected part of the correlation function. 
Here the spacetime integral with the subscript symbol $T$ means $\int_0^{\beta=1/T} d\tau \int d^3 x $, reflecting the currently employed 
imaginary time formalism. 
Following the literature~\cite{Hatsuda:1994pi}, the explicit formula for $\chi_\pi$ 
in the present NJL model reads 
\begin{equation}
	\chi_{\pi}=\frac{\Pi_{\pi}\left(0, 0\right)}{1-\left[2G-K\left \langle \bar{s}s \right\rangle\right]\Pi_{\pi}\left(0, 0\right)},
\end{equation}
with 
$\Pi_\pi(0,0) \equiv \Pi_\pi(\omega, \vec{p})$ being the polarization (correlation) function for the $\pi$ channel. 
Presently, we focus only on the neutral meson channel, so that 
$\Pi_\pi(0,0) $ is evaluated as 
\begin{equation}
	\begin{aligned}
\Pi_{\pi}\left(0, 0\right)=-i \cdot \sum_{f=u, d}\left( \int \frac{d^{4} p}{(2 \pi)^{4}} \operatorname{tr}\left[i\gamma_{5} S^{f}(p)     i\gamma_{5}S^{f}(p)\right]\right).
	\end{aligned}
	\label{pifunc}
\end{equation}

Similarly, we next evaluate the $\delta$ meson susceptibility, which is 
defined as 
\begin{equation}
	\chi_{\delta} 
	=\int_T d^4 x 
	\left[ 
	\langle ( \bar u(0)  u(0))( \bar u(x) u(x))\rangle_{\rm conn}
+ \langle ( \bar d(0)  d(0))( \bar d(x) d(x))\rangle_{\rm conn} 
\right]
\,. \label{chi-delta:def}
\end{equation}
The explicit formula for $\chi_\delta$ reads~\cite{Hatsuda:1994pi} 
\begin{equation}
	\chi_{\delta}=\frac{\Pi_{\delta}\left(0, 0\right)}{1-\left[2G+K\left \langle \bar{s}s \right\rangle\right]\Pi_{\delta}\left(0, 0\right))} \,. 
\label{deltafunc}
\end{equation}
Focusing on the neutral $\delta$ meson component, 
we find the corresponding polarization function in the $\delta^{0}$ meson channel: 
\begin{equation}
	\begin{aligned}
\Pi_{\delta}\left(0, 0\right)=-i \cdot \sum_{f=u, d}\left( \int \frac{d^{4} p}{(2 \pi)^{4}} \operatorname{tr}\left[ S^{f}(p)     S^{f}(p)\right]\right).
	\end{aligned}
\end{equation}
The more detailed expressions for $\Pi_\delta$ as well as $\Pi_\pi$ 
are presented in Appendix~A, which are useful for performing numerical analysis.

 It is crucial to note that $\chi_\pi$ is related to 
 the light quark condensates, through the chiral Ward-identity~\cite{GomezNicola:2016ssy,GomezNicola:2017bhm,Kawaguchi:2020qvg} as 
 \begin{align} 
  \langle \bar{u}u \rangle + \langle \bar{d}d \rangle  = - m_0 \chi_\pi
\,. \label{WI}
 \end{align} 
This is operative even at finite temperature~\cite{Kawaguchi:2020qvg}, 
and can also work even with a constant magnetic field, though it 
provides extra explicit-chiral/isospin breaking term. 
The recent lattice 2 + 1 flavor simulation~\cite{Ding:2020hxw} have also proved that 
when $m_u=m_d$, 
Eq.(\ref{WI}) indeed holds in the magnetic field for each of $u$ and $d$ terms 
at the operator level, 
so it should work also at any temperature~\footnote{
We thank Heng-Tong Ding for this information. 
}. 
Indeed, we have checked that Eq.(\ref{WI}) is satisfied with $\chi_\pi$ 
and light quark condensates.

\begin{figure}[t]
	\centering
	\includegraphics[width=0.50\textwidth]{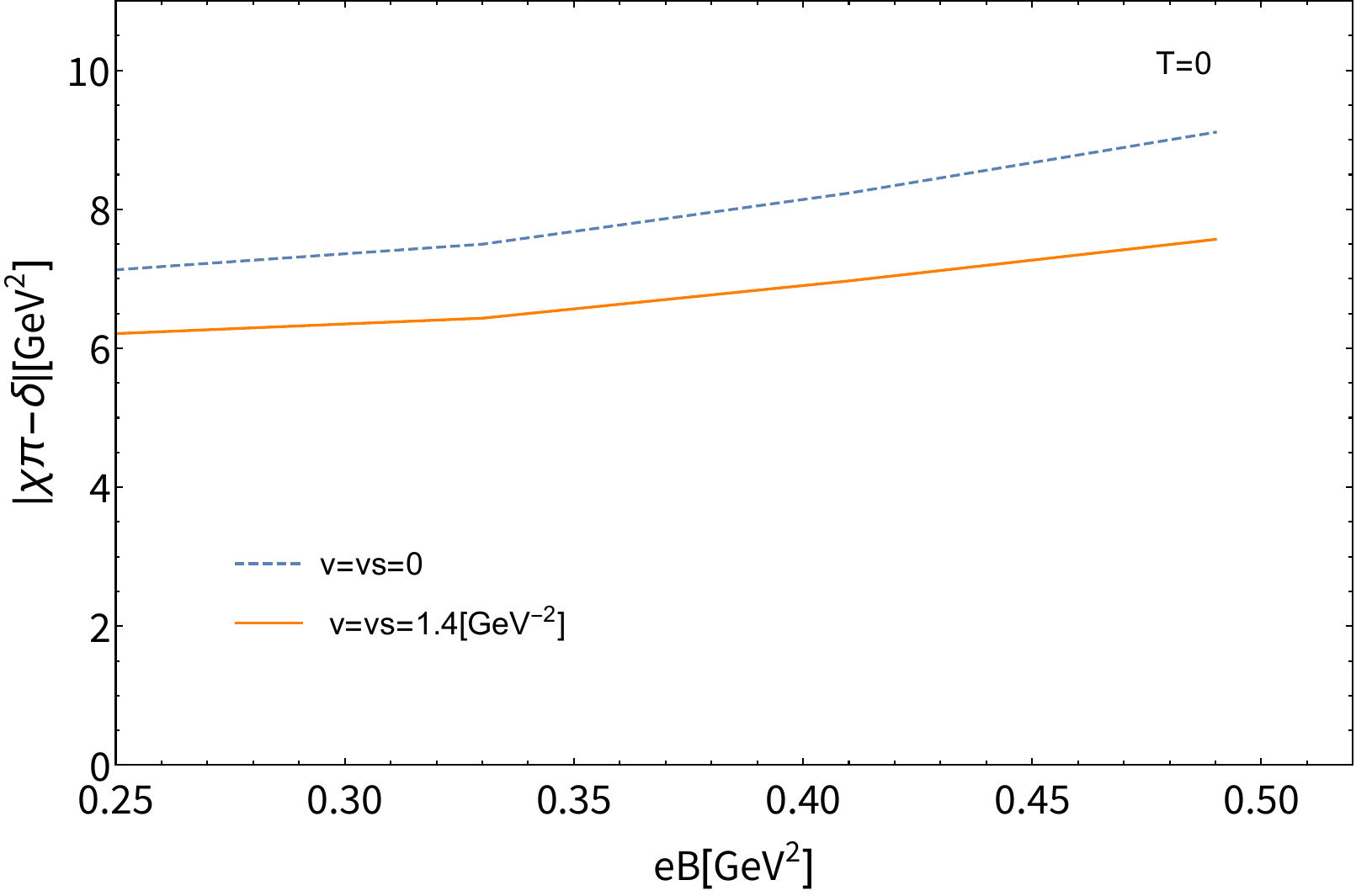}
	\caption{
	The magnitude of the axial susceptibility, $|\chi_{\pi-\delta}|$, versus the strength of $eB$, at zero temperature, with a couple of reference AMM values 
	inside and outside the ``Allowed" regime in Fig.~\ref{CIMC range11}. 
	Both cases realize the AMC. 
	The size of $eB$ has been bounded from below, at $eB \ge 0.25\,{\rm GeV}^2$, which is 
	to be consistent with quantitative agreement of the present NJL model on the 
	$eB$ dependence of the subtracted quark condensates at $T=0$ with those reported from 
	lattice QCD. For details, See the Summary and Discussion section, and Fig.~\ref{comparison}.  
	}
	\label{T equals zero11}
\end{figure}

Making difference of $\chi_\pi$ and $\chi_\delta$, 
we define the axial susceptibility as 
\begin{equation}
	\begin{aligned}
\chi_{\pi - \delta} \equiv \chi_{\pi}-\chi_{\delta}.
	\end{aligned}
\label{chi-pi-delta}
\end{equation}
This $\chi_{\pi- \delta}$ becomes zero, when the $U(1)_A$ symmetry is exact, 
because $\chi_{\pi} \leftrightarrow \chi_\delta$ by the $U(1)_A$ transformation, 
as is manifest in the definitions, Eq.(\ref{chi-pi:def}) and (\ref{chi-delta:def}). 
In Fig.~\ref{T equals zero11} we plot the magnitude of $\chi_{\pi -\delta}$ 
as a function of $eB$ at $T=0$, with the AMM values fixed inside (for $v=v_s=1.4\, {\rm GeV}^{-2}$ ) or outside (for $v=v_s=0$) the allowed regime in the ``phase" diagram, Fig.~\ref{CIMC range11}. 
We observe that $|\chi_{\pi-\delta}|$ gets larger as $eB$ grows, namely,   
the MC for the axial symmetry at $T = 0$.  
Realization of the AMC at $T=0$ is somewhat insensitive to the size of the AMMs, 
in sharp contrast to the case of the CMC at $T=0$.

\begin{figure}[t]
	\centering
	\includegraphics[width=0.50\textwidth]{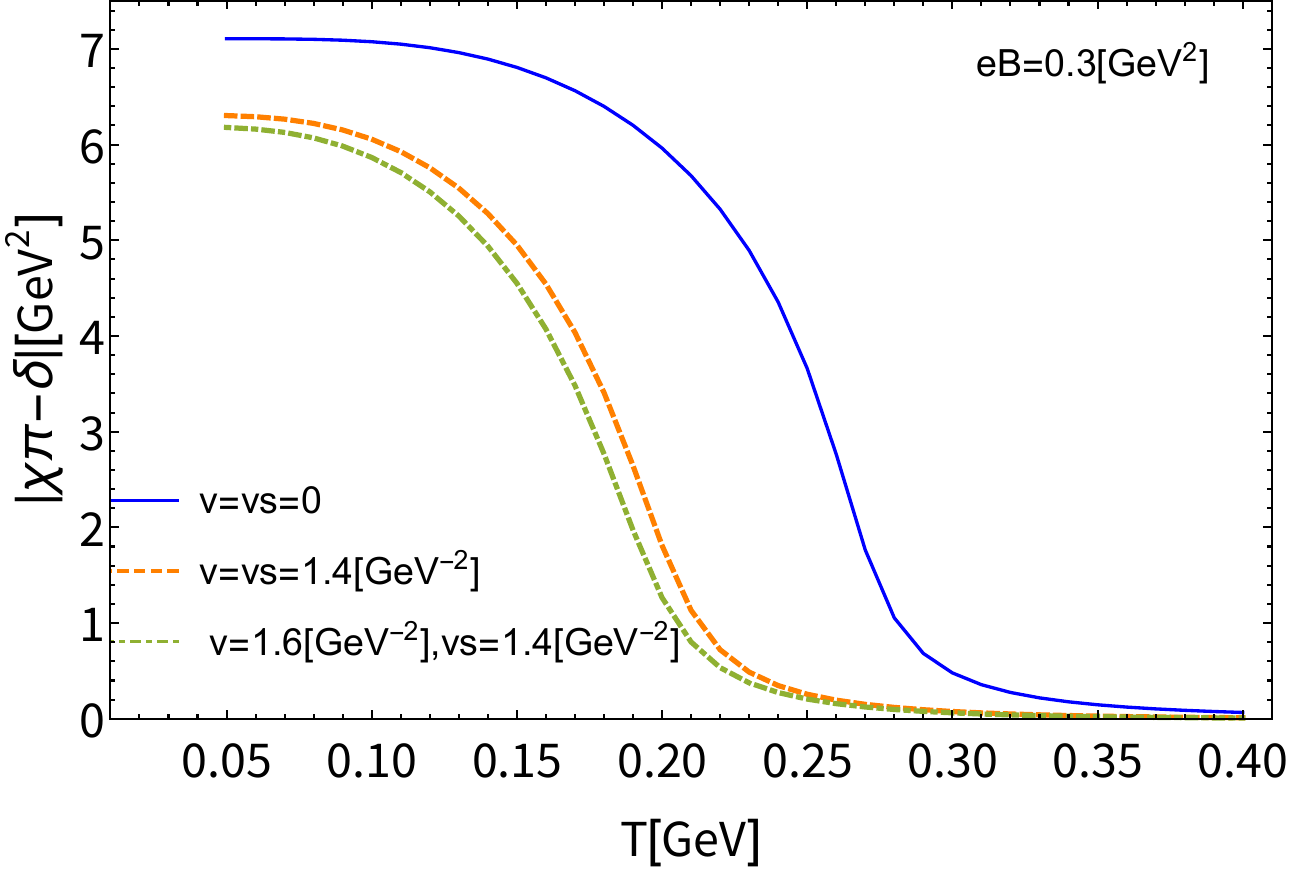}
	\caption{The AMM dependence on $|\chi_{\pi-\delta}|$ at finite temperature, with $eB$ fixed. }
	\label{chi with AMM and without 11}
\end{figure}

From Fig.~\ref{chi with AMM and without 11}, we also notice the trend of monotonic reduction for the magnitude by finite AMMs, with fixed $eB$, which is observed irrespective to $v$ or $v_s$.  
This is because all the AMMs play a destructive interference in $\chi_{\pi- \delta}$, 
against contributions from the current quark masses and 
the KMT-determinant $U(1)_A$ anomaly, to drive faster $U(1)_A$ restoration.  

This trend is observed even at finite temperature. 
See Fig.~\ref{chi with AMM and without 11}, 
which shows $|\chi_{\pi -\delta}|$ as a function of $T$, with $eB$ fixed and 
the sizes of the AMMs being flavor symmetrically ($v=v_s$) or asymmetrically ($v \neq v_s$) varied. 
Larger AMMs tend to reduce the magnitude of $|\chi_{\pi -\delta}|$ at any temperature, 
and the AMMs generically play a role of catalizer toward the $U(1)_A$ symmetry restoration.

\begin{figure}[t]
	\centering
	\includegraphics[width=0.50\textwidth]{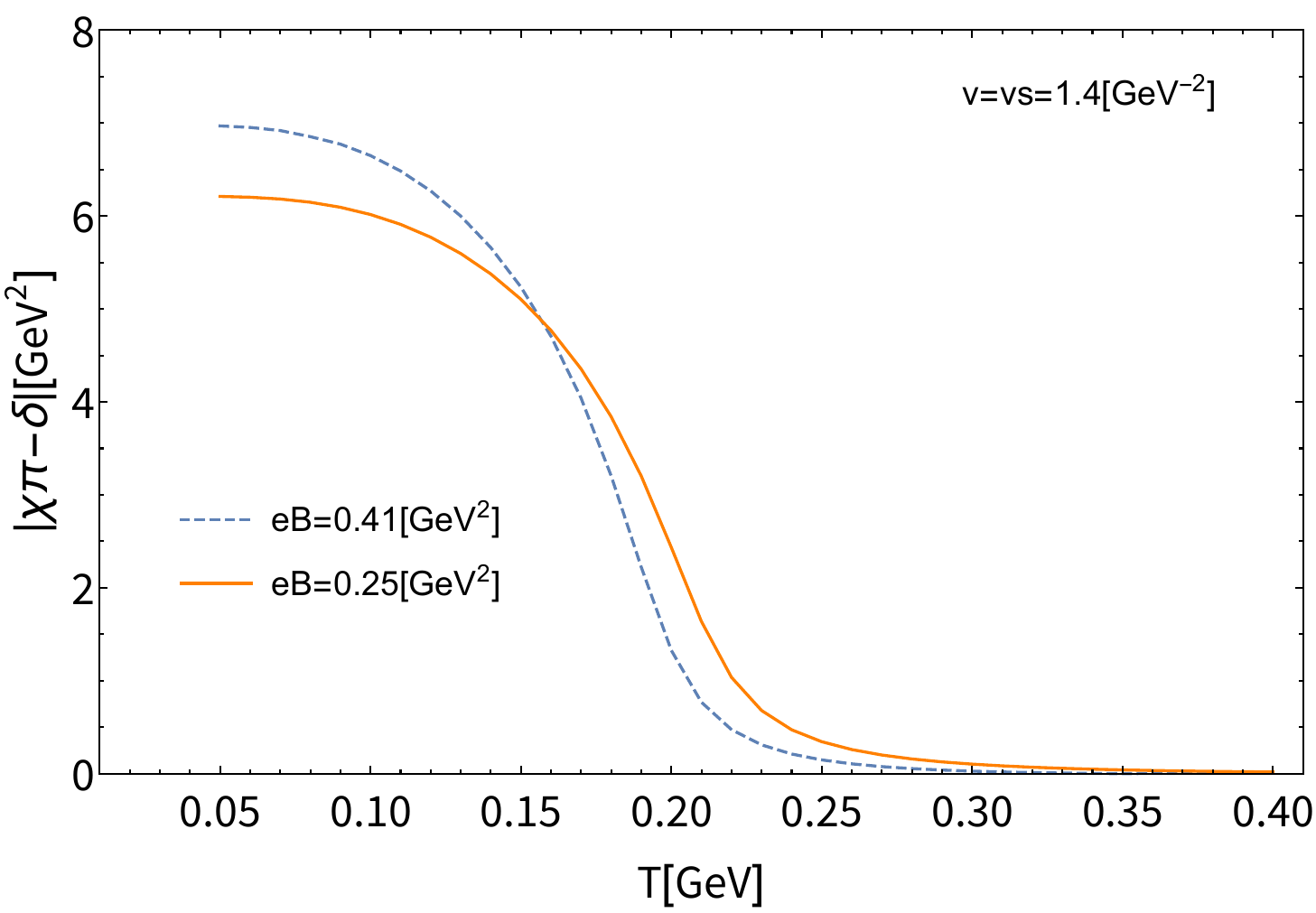}
	\caption{Plots of $|\chi_{\pi -\delta}|$ as a function of $T$, with $eB$ varied, at 
	a viable reference point for the AMMS $(v,v_s)$ in the ``phase" diagram, Fig.~\ref{CIMC range11}. The AIMC is observed. }
	\label{AIMC11}
\end{figure}
\begin{figure}[t]
	\centering
	\includegraphics[width=0.55\textwidth]{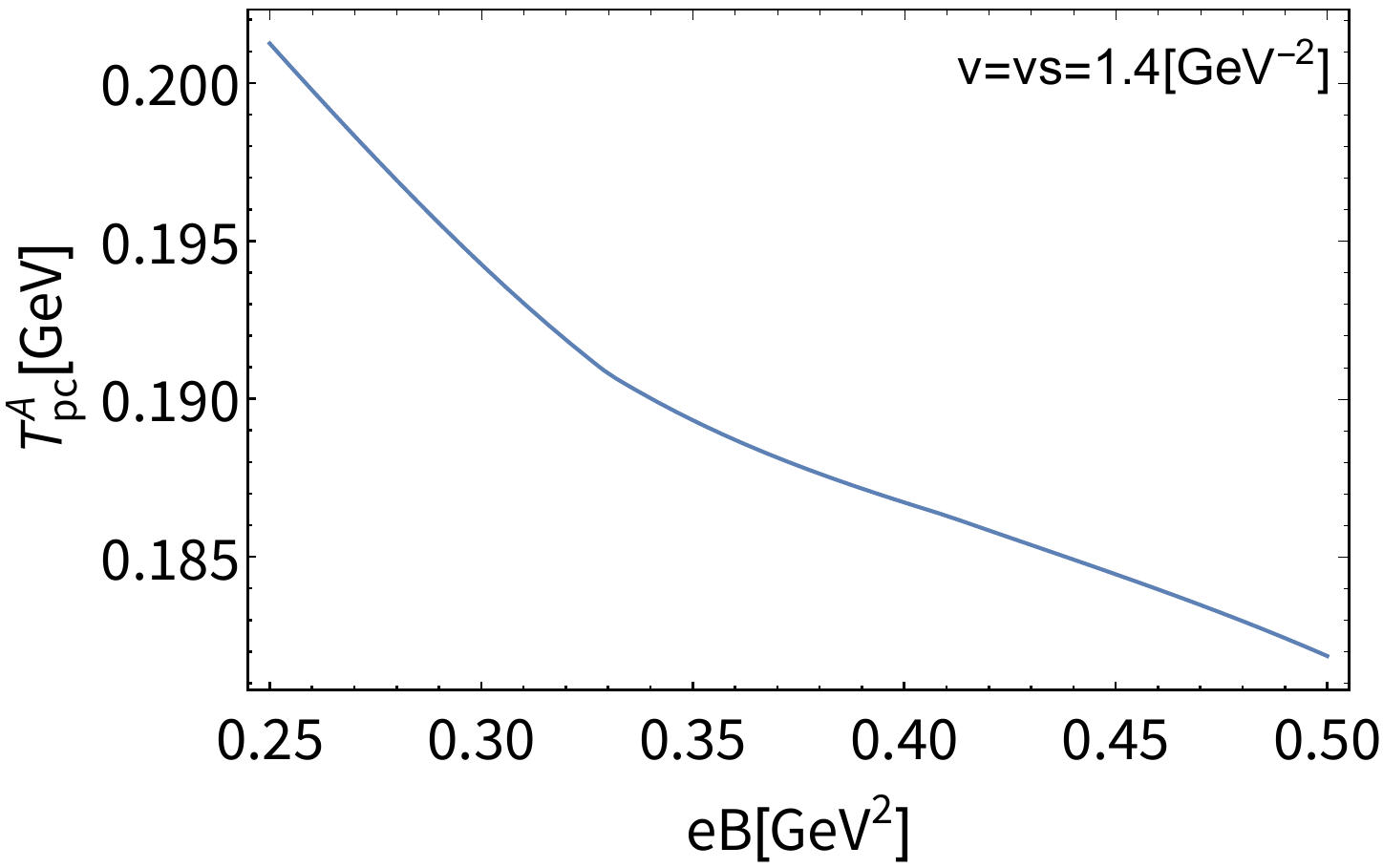} 
	\caption{
	The $eB$ scaling of the pseudo-critical temperature for $|\chi_{\pi-\delta}|$, 
	$T_{pc}^A$ defined in the text. 
	The AMM values have been set to the same  viable reference point as in Fig.~\ref{AIMC11}. The observed decreasing trend manifests presence of the AIMC. 
	The range of $eB$ has been restricted to be $\ge 0.25\, {\rm GeV}^2$, for the same reason as noted in the caption of Fig.~\ref{T equals zero11}. }
	\label{tac over tcc11}
\end{figure}

Figure~\ref{AIMC11} shows the $T$-dependence of $|\chi_{\pi- \delta}|$ with $eB$ varied, 
at a reference allowed point for the AMMs $(v, v_s)$ in Fig.~\ref{CIMC range11}. 
We see that $|\chi_{\pi - \delta}|$ starts to drop faster at higher temperatures, 
as $eB$ gets larger, while it develops with $eB$ at lower temperatures. 
This implies the IMC for the $U(1)_A$ symmetry, i.e., AIMC, 
in perfect analogy to the IMC for the chiral symemtry (CIMC). 
This AIMC can further be quantified by observing the $eB$ dependence on the pseudo-critical 
temperature, $T_{pc}^A$, which is defined as the inflection point of 
the $|\chi_{\pi- \delta}|$ curve with respect to $T$ as 
\begin{align} 
 \frac{\partial^2 |\chi_{\pi - \delta}| }{\partial T^2} \Bigg|_{T=T_{pc}^A} = 0
 \,. 
\end{align}
Figure~\ref{tac over tcc11} plots this $T_{pc}^A$ as a function of 
$eB$, at the same reference point for AMMS as in Fig.~\ref{AIMC11}. 
A monotonic decrease trend for $T_{pc}^A$ with growing $eB$ is indeed observed, 
so it manifests the AIMC, 
just like the case of the CIMC.


\begin{figure}[t]
	\centering
	\includegraphics[width=0.50\textwidth]{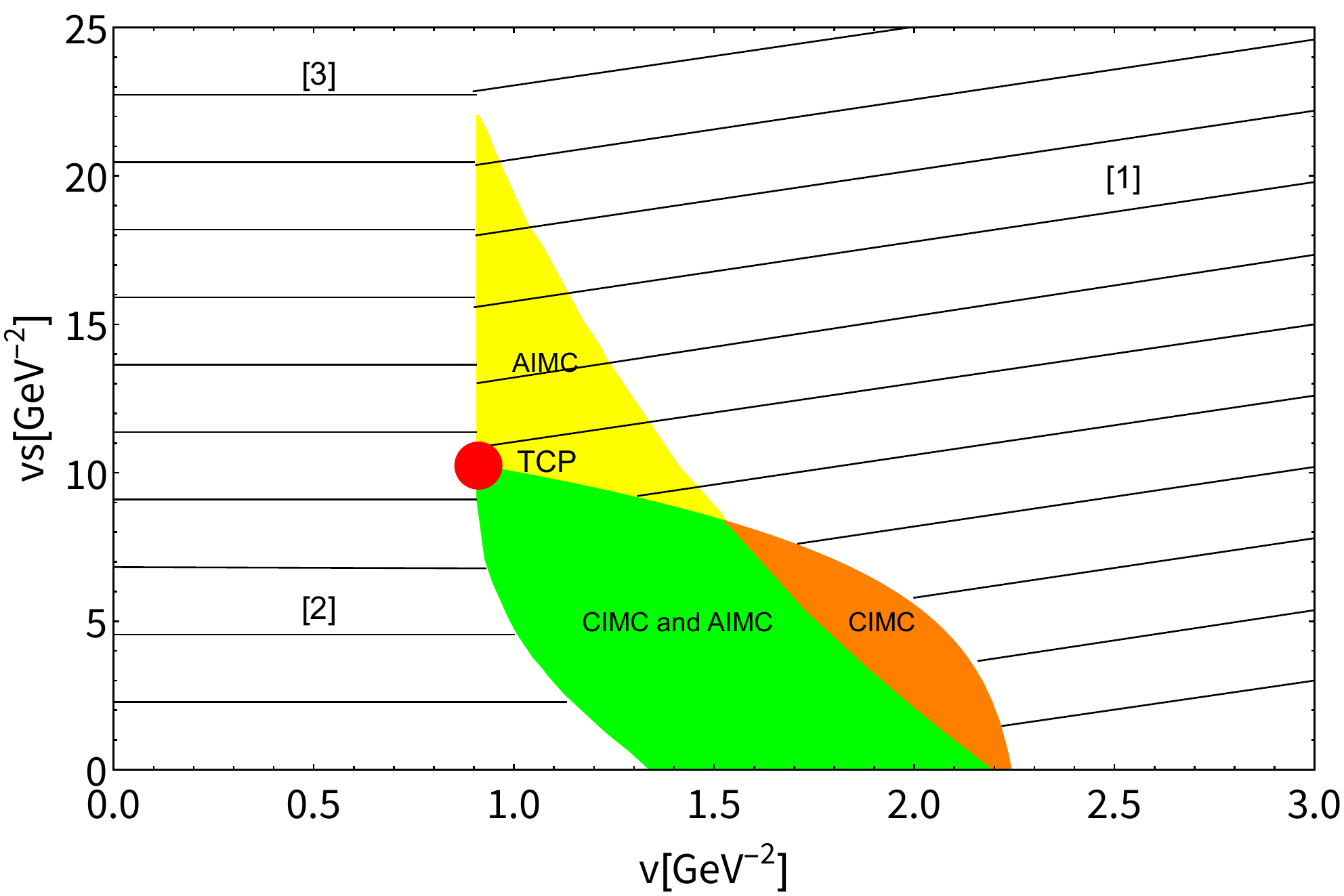}
	\caption{
	The ``phase" diagram extended from Fig.~\ref{CIMC range11} with 
	the property of the axial sector incorporated. In the same way as in Fig.~\ref{CIMC range11}, the ``phase" [1] is excluded by CIMC at $T$=0; 
	[2] is ruled out by CMC at higher temperatures; 
	[3] is disfavored because of CIMC at $T$ =0 and CMC at higher temperatures.  
	The TCP is placed at the same point as in Fig.~\ref{CIMC range11}.  
	The domain filled in yellow, which realizes the AIMC at higher temperatures, 
	almost overlaps with the ``Allowed" regime in Fig.~\ref{CIMC range11}, but 
	separates into two, to create the orange regime out of the original green ``Allowed" regime. The reason would be related to the calculability of the present analysis. 
	For more details, see the text. 
}
	\label{AIMC range11}
\end{figure}

Finally, in Fig.~\ref{AIMC range11} 
we show an extended ``phase" diagram on the AMM $(v, v_S)$ plane of Fig.~\ref{CIMC range11}, by incorporating 
the parameter space to realize the AIMC at high temperatures. 
The figure tells us that the AIMC at high temperature is necessarily realized 
when the desired CMC at lower $T$ and the CIMC at higher $T$ are present (filled in 
green), except for domains with a larger AMM for strange quark (in yellow) including the 
TCP (red blob), or with a larger AMM for up and down quarks (in orange).

The former discrepancy is due to the flavor-universal destructive contribution from the 
AMMs to $|\chi_{\pi -\delta}|$, hence even so large $v_s$ can still act as 
a destructive interference in $|\chi_{\pi -\delta}|$ against the other axial 
breaking contributions from the current quark mass and $U(1)_A$ anomaly. 
This feature is contrast to the $v_s$ sensitivity to the chiral symmetry, as emphasized above, for which the property of the $v_s$ interference changes in low or high temperatures.

The latter case would involve limitation of the present analysis 
based on the NJL description.  
The boundary separating the ``CIMC and AIMC" (in green) and ``CIMC" (in orange) domains 
has been created by the present calculability: going over the ``CIMC and AIMC" domain to the right 
(i.e. to a larger $v$ regime), 
$|\chi_{\pi-\delta}|$ actually starts to show non-monotoic damping scaling 
at higher $T$, which we would regard as unphysical or an artifact of the present NJL model as a low energy description of QCD. 
This lack of calculability has also affected the existence of  
a top endpoint of the AIMC regime (at $(v, v_s)\simeq (0.9, 22)$), 
and the right-side boundary curve in Fig.~\ref{AIMC range11}.

The left-side lower boundary curve (green part) is present because the AMC at higher temperatures 
is realized due to too small AMMs, which corresponds to the ``phase" [2], 
while the upper boundary curve (yellow part) has been created because of 
lack of calculability due to too large $v_s$, similarly to the aforementioned 
case with too large $v$.

Lattice simulations in the near future will clarify  the AIMC, and 
give a conclusive answer to 
whether the AIMC at high temperatures is necessarily realized when 
the CIMC at high temperatures is present.

\section{Summary and Discussion}

We found the AIMC at high temperatures, 
which can be driven by the 2 + 1 flavor-universal destructive interference against  
the current quark mass and the $U(1)_A$ anomaly in the axial susceptibility, 
when the CIMC is present. 
One possible candidate to realize this kind of 
destructive effects involves the AMMs of quarks, 
which, in the present paper, we have incorporated into 
an NJL model with 2 + 1 flavors, to observe emergence of 
the AIMC at the physical point. 
Measuring the AIMC  
would give a complementary probe of the CIMC (See Fig.~6), and  
imply a definite interpretation on how the IMC is realized: 
it is due to the destructive axial- and chiral-breaking driven in 
thermomagnetic QCD. 
Our findings are shortly testable by lattice simulations in the near future, 
by which the AMM values in the NJL model can be constrained and 
a part of the model parameter space will be excluded or probed.

Several comments are in order.

\begin{itemize} 

\item 

The presence of the AIMC  
manifests  the intrinsic and non-factorizable chiral-axial correlation, 
which has been supported from a recent rigorous proof by the lattice  study~\cite{Ding:2020xlj}.  
This correlation can be suspected also from the result in~\cite{Cui:2021bqf} 
at zero magnetic field, based on the same 
NJL model with 2 + 1 flavors as in the present paper, 
which is tightly constrained by single anomalous 
-chiral Ward identity involving the chiral and axial susceptibilities, 
together with the topological susceptibility.

\begin{figure}[t]
	\centering
	\includegraphics[width=0.45\textwidth]{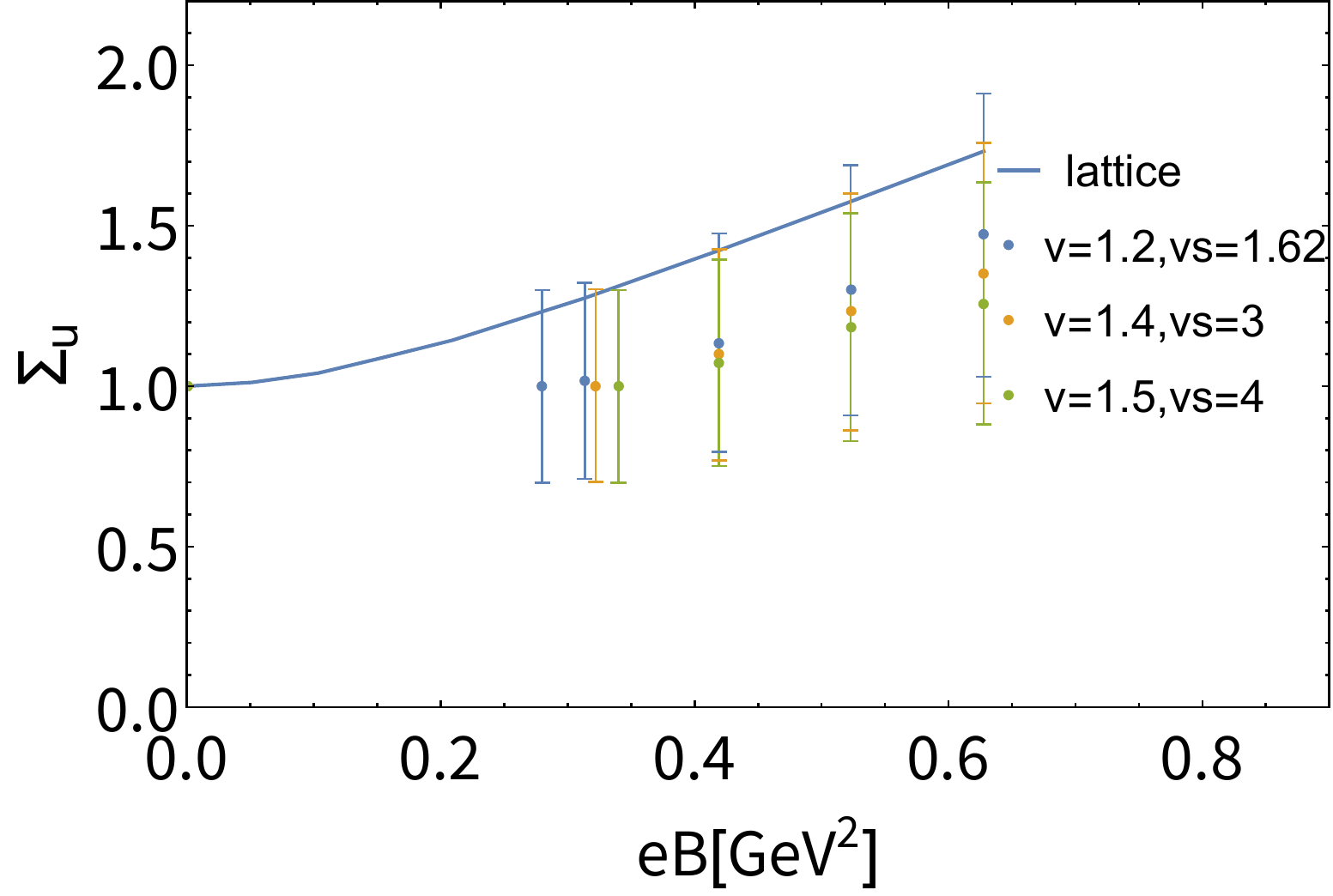}
	\includegraphics[width=0.45\textwidth]{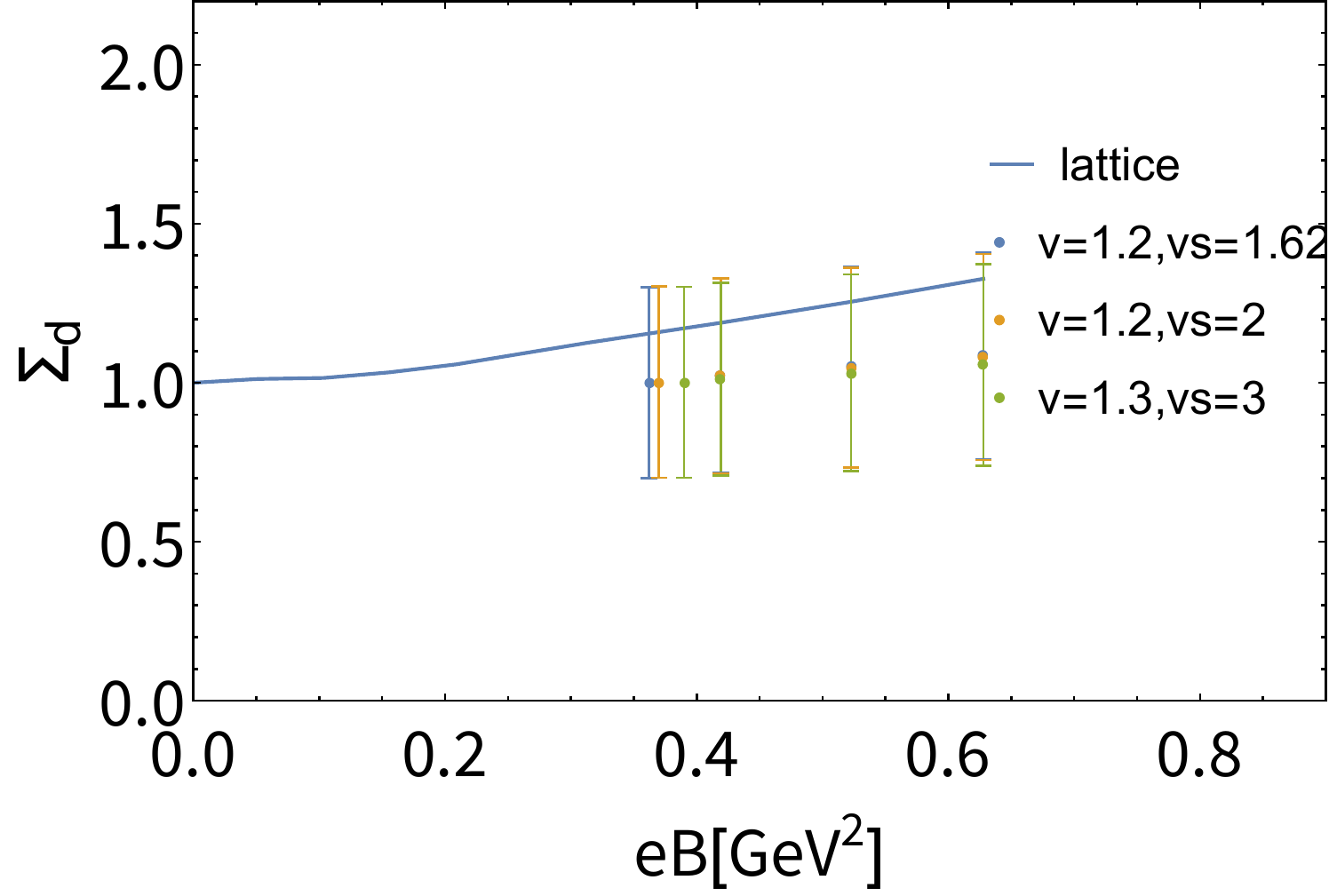}
	\caption{Comparison of $eB$ dependence on the 
	subtracted quark condensates $\Sigma_u$ (left panel) and $\Sigma_d$ (right panel)  at $T=0$, with the 2 +1 flavor lattice data (solid curve)~\cite{Ding:2020hxw}. Error bars along with the model yields 
	correspond to the 30\% uncertainty, described in the text. The values of $v$ and $v_s$ are read with unit of ${\rm GeV}^{-2}$. 
	}
	\label{comparison}
\end{figure}

\item 

Though both emergence of the CIMC and AIMC are 
qualitatively interpreted by presence of AMM 
parameters for quarks and its destructive interference against 
the current quark mass and axial anomaly effects, 
we may check its quantitative consistency of the presence of 
such sizable parameters, by evaluating the $eB$ dependence 
of the subtracted light-quark condensate $\Sigma_{f} \equiv 
1 + \frac{2 m_{0f}}{m_\pi^2 f_\pi^2} (\langle  - \bar{q}_f q_f \rangle (eB \neq 0) - \langle - \bar{q}_f q_f\rangle (eB=0))$ at $T=0$ for $q_f=u,d$, 
in comparison 
with the lattice 
data in~\cite{Ding:2020hxw}. 
Here we take into account a 30\% uncertainty, which corresponds to 
the order of subleading corrections in the large $N_c$ expansion, 
on that the present NJL-model analysis has been based. 
This comparison would further constrain the parameter space $(v, v_s)$ in Fig.4  and 
the applicable size of the strong magnetic field strength: 
$1.2 \lesssim v \lesssim 1.5$, $v_s \lesssim 4$, in unit of ${\rm GeV}^{-2}$, 
and $0.25 \lesssim eB_{\rm min} \lesssim 0.34$ in unit of ${\rm GeV}^{2}$. 
Thus too large $v$ and $v_s$ would 
drive $\Sigma_{u,d} <1$, hence would be disfavored because of 
spoiling the magnetic catalysis at $T=0$, which also places the lower bound 
on the strength of $eB$. 
This is due to too large destructive 
interference from the AMM contributions 
to the quark condensate. 
When this quantitative constraint is combined with 
the allowed regime in the ``phase " diagram, Fig.~7, 
we see that the AIMC may necessarily 
take place at the same time the CIMC is present.

More precise $eB$ and $T$ dependence of the chiral and axial susceptibilities 
could be determined by fitting all the model parameters to available 
lattice data, as 
has been discussed in the literature~\cite{Endrodi:2019whh} 
in light of the CIMC. 
Then one could more quantitatively discuss whether 
the AIMC can take place 
at the same time the chiral one does, though 
presence of the destructive interference would be obscure. 
This may be worth performing elsewhere.

\item 
 
The AIMC may provide a hint to reveal whether   
in a sense of early thermomagnetic universe, 
the remnant of the $U(1)_A$ breaking in the origin of mass 
might be comparable with what the chiral breaking leaves, 
in contrast to the pure-thermal QCD in which the former might highly dominate~\cite{Aoki:2021qws}.

\item 

Confirmation of the AIMC 
by lattice QCD simulations for 
the AIMC at high temperatures 
might give evidence of the destructive chiral and axial breaking 
in the chiral and the axial susceptibilities, 
and effective chiral models without such destructive interference leading to 
both the CIMC and AIMC would be excluded.

\item 

Possible prospected studies along this AIMC also include 
correlation between the dual CIMC and AIMC and 
the topological susceptibility: as briefly aforementioned, 
in the case without magnetic fields, 
the susceptibility differences for 
the chiral and axial partners are firmly linked to the topological 
susceptibility, through the anomalous-chiral Ward  identities~\cite{GomezNicola:2016ssy,GomezNicola:2017bhm,Kawaguchi:2020qvg}, 
and can form what is called the QCD trilemma~\cite{Cui:2021bqf}. 
It would be noteworthy to look into the magnetic dependence on the 
topological susceptibility, when a strong magnetic field is applied, 
and its sensitivity to the emergence of the chiral and axial inverse magnetic catalyses, through 
the chiral Ward identities~\footnote{
In Ref.~\cite{Sabir:2020sje} a possible correlation between 
the CIMC and an IMC for the topological susceptibility has been discussed 
based on a different type of NJL-like description with 2 flavors, 
where the axial susceptibility like $\chi_{\pi -\delta}$ is not addressed, however. 
Note that the axial breaking, i.e., nonzero $\chi_{\pi-\delta}$ is not always equal to 
the topological susceptibility $\chi_{\rm top}$, as has recently been 
clarified in~\cite{Cui:2021bqf} by deeply studying 
the anomalous chiral-Ward identity. 
More precisely and rigorously, it involves the chiral breaking as well, 
and becomes equal to $\chi_{\rm top}$ only when the chiral 
 symmetry is completely restored (at $T \to \infty)$. 
Therefore, the $T$-dependence and quark mass dependence of 
$\chi_{\pi-\delta}$ cannot simply follow that of $\chi_{\rm top}$. 
}.

\item 

Some cosmological implications to QCD axion coupled to 
thermomagnetic QCD could also be derived: if the topological 
susceptibility could drop faster around the chiral crossover regime, 
due to the strong magnetic field along with the CIMC and AIMC, 
the axion mass (with fixed axion decay constant) 
could also get smaller, implying a significant modification 
of estimate on the thermal relic abundance of axion as a dark matter candidate.

\item 

Other possible application of the AIMC may be related to  
a stable neutral $a_0$ meson around the chiral crossover regime: 
note first that the susceptibilities scale with the associated meson masses like $\sim 1/m_{\rm mesons}^2$, hence the degeneracy in the meson masses actually 
signals the (effective) restoration of the associated symmetry. 
Since we expect $m_{\pi^0} \simeq m_{(\delta^0 = a_0^0)}$ due to the AIMC, 
the dominant decay channel $a_0^0 \to \pi^0 + \eta^0$ will almost be closed 
(no matter how the $\eta^0$ mass gets changed from the vacuum value around the chiral crossover regime), 
so that the neutral $a_0$ meson can be somewhat long-lived. This might give phenomenological 
impact on meson physics 
relevant to heavy ion collision experiments. 

\end{itemize}

Exploring those interesting issues are to be left, and pursed elsewhere.

\section*{Acknowledgements}

\vspace{15pt}
We are grateful to 
Heng-Tong Ding, Chowdhury Aminul Islam,  
Mamiya Kawaguchi, Sheng-Tai Li, Akio Tomiya, and Lang Yu  
for useful comments. 
This work was supported in part by the National Science Foundation of China (NSFC) under Grant No.11747308, 11975108, 12047569, 
and the Seeds Funding of Jilin University (S.M.).

\appendix

\section{Useful formulas for polarization functions $\Pi_\pi$ and $\Pi_\delta$} 
Substituting the quark propagator under the magnetic field in Eq.(\ref{Sf}), which includes the AMM terms,  into the polarization functions, we work on 
the Dirac trace quantities.  
For the pion polarization function in Eq.(\ref{pifunc}), we have 
\begin{equation}
\begin{aligned}
&\operatorname{Tr}\left[\gamma_{5} D_{n}\left(p_{\|}, p_{\perp}\right) F_{n}\left(p_{\|}, p_{\perp}\right) \gamma_{5} D_{m}\left(p_{\|}, p_{\perp}\right) F_{m}\left(p_{\|}, p_{\perp}\right)\right] \\&=-8\left[f_{1}(n) f_{1}(m)-f_{2}^{2} \cdot p_{\|}^{2}\right]\left[\left(L_{n} L_{m}+L_{n-1} L_{m-1}\right) \cdot\left[p_{\|}^{2}-\left(M_{f}^{2}+\kappa_{f}^{2} q_{f}^{2} B^{2}\right)\right]\right. \\
&\left.-2 \, {\rm sign}\left(q_{f} B\right) M_{f} \kappa_{f} q_{f} B\left(L_{n} L_{m}-L_{n-1} L_{m-1}\right)-8 p_{\perp}^{2} L_{n-1}^{1} L_{m-1}^{1}\right],
\end{aligned}
\end{equation}
where we have omitted the arguments for the generalized  Laguerre polynomials, 
and defined 
\begin{equation}
\begin{aligned}
f_{1}(n) & \equiv p_{\|}^{2}+\left(\kappa_{f} q_{f} B\right)^{2}-M_{f}^{2}-2 n\left|q_{f} B\right| ,\\
f_{2} & \equiv -2 \kappa_{f} q_{f} B.
\end{aligned}
\end{equation}

In a similar way, for the delta meson polarization function in Eq.(\ref{deltafunc}), we have 
\begin{equation}
\begin{aligned}
&\operatorname{Tr}\left[D_{n}\left(p_{\|}, p_{\perp}\right) F_{n}\left(p_{\|}, p_{\perp}\right) D_{m}\left(p_{\|}, p_{\perp}\right) F_{m}\left(p_{\|}, p_{\perp}\right)\right]\\
&=8\left[f_{1}(n) f_{1}(m)+f_{2}^{2} \cdot p_{\|}^{2}\right]\left[\left(L_{n} L_{m}+L_{n-1} L_{m-1}\right) \cdot\left[p_{\|}^{2}+\left(M_{f}^{2}+\kappa_{f}^{2} q_{f}^{2} B^{2}\right)\right]\right. \\
&\left.+2 \, {\rm sign}\left(q_{f} B\right) M_{f} \kappa_{f} q_{f} B\left(L_{n} L_{m}-L_{n-1} L_{m-1}\right)-8 p_{\perp}^{2} L_{n-1}^{1} L_{m-1}^{1}\right] \\
&+16\left(f_{1}(n)+f_{1}(m)\right) f_{2} p_{\|}^{2} M_{f} \, {\rm sign}\left(q_{f} B\right)\left(L_{n} L_{m}-L_{n-1} L_{m-1}\right) \\
&-8 p_{\|}^{2}\left(f_{1}(n)+f_{1}(m)\right) f_{2}^{2}\left(L_{n} L_{m}+L_{n-1} L_{m-1}\right),
\end{aligned}
\end{equation}
where
\begin{equation}
\begin{aligned}
f_{1}(n) &=p_{\|}^{2}+\left(\kappa_{f} q_{f} B\right)^{2}-M_{f}^{2}-2 n\left|q_{f} B\right| , \\
f_{2} &=-2 \kappa_{f} q_{f} B .
\end{aligned}
\end{equation}

Plugging those into Eq.(\ref{chi-pi-delta}) with Eqs.(\ref{chi-pi:def}) and (\ref{chi-delta:def}), and working on momentum integration with respect to $p_{\perp}$, 
and summing over Matsubara frequencies, 
we are then ready to evaluate the axial susceptibility numerically. 
Through this procedure, a couple of the results have been presented in the main text.

\end{document}